\documentclass{article}
\usepackage{graphicx} % Required for inserting images
\usepackage[a4paper,top=3cm,bottom=2cm,left=3cm,right=3cm,marginparwidth=1.75cm]{geometry}

\usepackage{amsmath}
\usepackage{amssymb}
\usepackage[colorinlistoftodos]{todonotes}
\usepackage[colorlinks=true, allcolors=blue]
{hyperref}
\usepackage{lineno}
\usepackage{setspace}
\usepackage[numbers]{natbib}

\title{From Exponential to Polynomial Complexity: Efficient Permutation Counting with Subword Constraints}
\author{Martin Mathew, Javier Noda}

\date{September 2024}

\begin{document}

\maketitle

\section{Abstract}

Counting distinct permutations with replacement, especially when involving multiple subwords, is a longstanding challenge in combinatorial analysis, with critical applications in cryptography, bioinformatics, and statistical modeling. This paper introduces a novel framework that presents closed-form formulas for calculating distinct permutations with replacement, fundamentally reducing the time complexity from exponential to linear relative to the sequence length for single-subword calculations. We then extend our foundational formula to handle multiple subwords through the development of an additional formula. Unlike traditional methods relying on brute-force enumeration or recursive algorithms, our approach leverages novel combinatorial constructs and advanced mathematical techniques to achieve unprecedented efficiency. This comprehensive advancement in reducing computational complexity not only simplifies permutation counting but also establishes a new benchmark for scalability and versatility. We also demonstrate the practical utility of our formulas through diverse applications, including the simultaneous identification of multiple genetic motifs in DNA sequences and complex pattern analysis in cryptographic systems, using a computer program that runs the proposed formulae.

\section{Introduction}
    
Counting distinct permutations with replacement is a fundamental problem in combinatorial analysis, with critical applications in cryptography, bioinformatics, and statistical physics. While classical permutation counting without replacement has well-established solutions, the inclusion of replacement and the presence of repeated substructures introduce significant complexity. Real-world combinatorial problems often involve scenarios where multiple subwords appear simultaneously within sequences, which introduces significant positioning considerations. The computational challenge lies in efficiently determining the count of distinct permutations for sequences of arbitrary length, particularly when the number of possible sequences grows exponentially with increasing length.

Existing brute-force enumeration and recursive methods quickly become infeasible for large-scale problems, highlighting the need for scalable, general solutions. To address this, we introduce a novel framework that significantly reduces the computational complexity of permutation counting. Building on a foundational formula for single subwords without self-intersections, we further develop an aggregated formula, an extension that enables efficient calculation of distinct permutations involving multiple subwords. This extension significantly reduces time complexity relative to the sequence length, a substantial improvement over traditional approaches.

By leveraging advanced combinatorial techniques and introducing innovative constructs, our framework simplifies permutation counting and broadens its applicability to previously computationally infeasible combinatorial structures. The results presented herein establish a new benchmark for scalability and versatility, enabling large-scale analysis in fields requiring extensive permutation counting. In the sections that follow, we provide necessary background, derive the formulas, and demonstrate their utility through theoretical insights, comparative analyses, and practical applications, concluding with a discussion of future research directions

\section{Methodology and Derivation of Formulae}

\subsection{Single Subword}

\textbf{Notation.} We follow the notation described in \textit{Enumerative Combinatorics: Volume I} by Richard P. Stanley \cite{stanley2011}. Let \( \mathbb{N} \) denote the set of natural numbers. Let \( Q \) be an alphabet with \( |Q| = q \) distinct symbols. For a word \( W \) over the alphabet \( Q \), let \( \text{CT}(W) \) denote the multiset of all contiguous subwords of \( W \). Let \( t \in \mathbb{N} \) denote the length of words in consideration, \( T = Q^t \) the set of all words of length \( t \), and \( S = \{ \text{CT}(I) : I \in T \} \) the set of all multisets of contiguous subwords of words in \( T \). The multiplicity of an element \( l \) in a multiset is denoted by \( \nu(l) \).

\textbf{Definition 1.} An \textit{unoccupied position} in a word \( R \) is a position that is not part of any occurrence of a given subword \( A \).

\textbf{Definition 2.} A \textit{relative position} is defined as the number of copies of $A$ that precedes an unoccupied position in the word. 

\textbf{Definition 3.} A word $W$ \emph{self-intersects} if an end segment is the same as an initial segment and the segments are neither empty, nor the word $W$ itself.

\textbf{Lemma 4} Let: $j, m, n \in 	\mathbb{N}.$ Let $
k \in \mathbb{N}$. Let $ k \leq n.$

Then:
\begin{linenomath*}
    \begin{align}
        \sum_{i = j}^{m}(-1)^i\binom{n}{k-i} = (-1)^j\binom{n-1}{k-j}+ (-1)^m\binom{n-1}{k-m-1}  \newline
    \end{align}
\end{linenomath*}

\textit{Proof} Using the simplification for the alternating row sum \cite{plaza2016}:

\begin{linenomath*}
    \begin{align}
\sum_{i = j}^{m}(-1)^i\binom{n}{i} = (-1)^j\binom{n-1}{j-1}+(-1)^m\binom{n-1}{m}
    \end{align}
\end{linenomath*}

An integer \textit{k} is added to the bottom of the binomial coefficient, thus negating \textit{i}; doing so reverses the alternating sum with an offset of \textit{k} as such: 

\begin{linenomath*}
    \begin{align}
\sum_{i = j}^{m}(-1)^i\binom{n}{k-i} = (-1)^j\binom{n-1}{k-j}+ (-1)^m\binom{n-1}{k-m-1}\newline
   \end{align}
\end{linenomath*}

This changes the solution by negating both \textit{j} and \textit{n} and switching negative one from the \textit{j} term to the \textit{n} term and adding \textit{k} to both bottoms as an offset.

\textit{Qed.} \newline

\textbf{Theorem 5.} Assume $A$ is a word in $Q$ that does not self-intersect. Let $|A| = a.$ Let $x \in \mathbb{N}.$ Let $Y = \{ I  \in  S \ | \ \nu(A) = x\}.$  

Then:

 \begin{linenomath*}
    \begin{align}
|Y|=\sum_{i=x}^{\lfloor \frac{t}{a} \rfloor}  (-1)^{i+1}q^{t-ai}\Bigg(\binom{i+1}{t-ai}\Bigg)\binom{i}{i-x}
    \end{align}
\end{linenomath*}

$Proof.$ The summation in line 4 must end at $\lfloor\frac{t}{a}\rfloor$ because this is the maximum number of copies of $A$ that can fit into a word of length $t$. Each copy of $A$ occupies a number of positions, so multiplying $a$ by the index $i$ gives the total number of positions occupied by the $i$ copies of $A$. Therefore, the number of unoccupied positions is $(t-ia)$. The term $q^{t-ia}$ represents the number of ways to fill the unoccupied positions with letters from the alphabet $Q$. This accounts for all distinct permutations of the unoccupied positions. \newline \par

Since there are $i$ copies of $A$, an unoccupied position is preceded by any number of copies of $A$ from 0 to $i$, resulting in $i+1$ possible relative positions. The objective is to find how many ways there are to assign each of the $t-ia$ unoccupied positions to one of the $i+1$ relative positions. This can be stated as:
\begin{linenomath*}
    \begin{align}
\Bigg(\binom{i+1}{t-ia}\Bigg) \newline
    \end{align}
\end{linenomath*}

When the initial value of $x$ is even, the alternating term $(-1)^{i+1}$ is adjusted to $(-1)^{i}$. This ensures that the first iteration always yields a positive value. 

\textbf{Lemma 6.} Let $n_a,n_d, z, b \in \mathbb{N}.$ Let $ k_a \in \mathbb{N}. $ Let $k_a \leq n_a.$ Let $j \in \mathbb{N}.$ Let $j \leq n_d.$ \newline

Then:

    \begin{linenomath*}
    \begin{align}
    \sum_{i = z}^{b}(-1)^i\binom{n_a}{k_a - i}\binom{n_d+i}{i} = (-1)^z\sum_{i =1}^{j}\binom{n_a - i}{k_a - z}\binom{n_d +z - i}{z - i} 
        \end{align}
\end{linenomath*}
    \[+ \sum_{i = z}^{b}(-1)^i\binom{n_a - j}{k_a - i}\binom{n_d - j + i}{i} \]
    \[+ (-1)^b\sum_{i = 1}^{j}\binom{n_a - i}{k_a - b - i}\binom{n_d + b}{b}\]

$Proof.$ For the purposes of this Lemma, $(-1)^x$, where $x$ can be any variable, will not be considered a term. In this proof, the first combination term,$\binom{n_a}{k_a - i}$, will be referred to as the across-sum; the second combination term, $\binom{n_d+i}{i}$ will be referred to as the diagonal. Using Pascal’s Identity \cite{brualdi2010} of the binomial coefficient yields the recursive formula for the across-sum and the diagonal as such: 

 \begin{linenomath*}
    \begin{align}
    \sum_{i = z}^{b} (-1)^i(\binom{n_a - 1}{k_a - i} + \binom{n_a -1}{k_a - i -1})(\binom{n_d - 1 +i}{i}+\binom{n_d - 1 + i}{i - 1})
    \end{align}
\end{linenomath*}

Choose an integer $c$ such that $z<c<b$. This means that $c-1$ and $c+1$ are both in the interval of $z \leq c-1,c+1 \leq b$. Input the $c$ value into the recursive diagonal: 

 \begin{linenomath*}
    \begin{align}
    (-1)^c(\binom{n_d - 1 + c}{c} + \binom{n_d - 1 + c}{c - 1})
    \end{align}
\end{linenomath*}

 $c-1$ is inputted for $c$ into the same function in line 8:

 \begin{linenomath*}
    \begin{align}
    (-1)^{c-1}(\binom{n_d - 2 + c}{c-1} + \binom{n_d - 2 + c}{c - 2})
    \end{align}
\end{linenomath*}

$c-1$ is also inputted into the regular diagonal function for $i$ in line 6: 

 \begin{linenomath*}
    \begin{align}
    \binom{n_d+c-1}{c-1}
    \end{align}
\end{linenomath*}

Upon slight rearrangements, the function in line 10 is equal to the right term of the recursive function in line 8. Thus, in a diagonal, one term of the diagonal plus a new term is used to create the next part of the diagonal. It is possible to edit the recursive function of $c$ to be: 

 \begin{linenomath*}
    \begin{align}
    (-1)^c(\binom{n_d - 1 + c}{c} + \binom{n_d - 2 + c}{c-1} + \binom{n_d - 2 + c}{c - 2})
    \end{align}
\end{linenomath*}

The function at point $c$ in line 7 is simplified in relation to its neighbors as such:

 \begin{linenomath*}
    \begin{align}
    (-1)^c\binom{n_d-1+c}{c}\binom{n_a - 1}{k_a - c}
    \end{align}
\end{linenomath*}

If $c$ meets the criteria that $z<c<b$, and if the boundary cases are excluded, a new sum that contains the function in line 12 is created if $c$ is replaced with the index of the sum, $i$, as such:

 \begin{linenomath*}
    \begin{align}
    \sum_{i = z + 1}^{b-1}(-1)^i\binom{n_a - 1}{k_a - i}\binom{n_d - 1 + i}{i}
    \end{align}
\end{linenomath*}

There are boundary cases when $c=z$ and when $c=b$. The boundary cases will then multiply out. First consider when $c=z$: 

 \begin{linenomath*}
    \begin{align}
    (-1)^z\binom{n_a - 1}{k_a - z}\binom{n_d + z - 1}{z} + (-1)^z\binom{n_a - 1}{k_a -z}\binom{n_d +z - 1}{z-1}
    \end{align}
\end{linenomath*}

The left-most term of line 14 is the same as the earlier sum in line 13 with $i=z$. Thus, this is added back into the sum in line 13 by changing the bounds of the function. \par

For $c=b$, the function in line 7 simplifies as such:

 \begin{linenomath*}
    \begin{align}
    (-1)^b\binom{n_a - 1}{k_a - b}\binom{n_d + b - 1}{b} + (-1)^b\binom{n_a - 1}{k_a - b-1}\binom{n_d +b}{b}
    \end{align}
\end{linenomath*}

Note that the diagonal for the right term of line 15 is the sum of all the recursive combinations at the end. As before, the left term of $c=b$  in line 15 is joined with the sum in line 13 by changing the end bound to include $b$:

 \begin{linenomath*}
    \begin{align}
    (-1)^p\binom{n_a - 1}{k_a - z}\binom{n_d +z - 1}{z - 1}
    + \sum_{i = z}^{b}(-1)^i\binom{n_a - 1}{k_a - i}\binom{n_d - 1 + i}{i}
                \end{align}
\end{linenomath*}
\[+ (-1)^b\binom{n_a - 1}{k_a - b - 1}\binom{n_d + b}{b}\]

The recursive formula is applied to the middle term in line 16 $j$ times and simplified to: 

 \begin{linenomath*}
    \begin{align}
    (-1)^z\sum_{i =1}^{j}\binom{n_a - i}{k_a - z}\binom{n_d +z - i}{z - i}
        \end{align}
\end{linenomath*}
    \[+ \sum_{i = z}^{b}(-1)^i\binom{n_a - j}{k_a - i}\binom{n_d - j + i}{i}\] \[+ (-1)^b\sum_{i = 1}^{j}\binom{n_a - i}{k_a - b - i}\binom{n_d + b}{b}\]

$Qed.$ \newline

\textbf{Lemma 7}. Let $b=k_a=\lfloor\frac{t}{a}\rfloor -x.$ Let $ \ n_a=\lfloor\frac{t}{a}\rfloor.$ Let $ n_d=j=x.$ Let $ z=0.$ \newline

Let:

 \begin{linenomath*}
    \begin{align}
    \alpha = \sum_{i = z}^{b}(-1)^i\binom{n_a}{k_a - i}\binom{n_d +i}{i}
    \end{align}
\end{linenomath*}

When: $\lfloor\frac{t}{a}\rfloor = x$, then $\alpha = 1$; else $\alpha = 0$. \newline

\textit{Proof.} If $\lfloor\frac{t}{a}\rfloor = x$, then $\lfloor\frac{t}{a}\rfloor - x = 0$. Since $b = z  $ in line 18, then the terms of the summation only need to be calculated once with $i$ set to $0$:

\begin{linenomath*}
    \begin{align}
    \alpha = \binom{\lfloor\frac{t}{a}\rfloor}{0}\binom{x}{0} =\  1
    \end{align}
\end{linenomath*}

Now consider when $\lfloor\frac{t}{a}\rfloor < x$. This indicates that $b < 0$ and $b < z$. Thus, the summation will result in 0.

Finally, assume $\lfloor\frac{t}{a}\rfloor > x$. Then, to simplify the $\alpha$ function, use Lemma 6 and input the variables defined to produce:

 \begin{linenomath*}
    \begin{align}
    \sum_{i = 0}^{\lfloor\frac{t}{a}\rfloor - x}(-1)^i\binom{\lfloor\frac{t}{a}\rfloor - x}{\lfloor\frac{t}{a}\rfloor -x - i}
    \end{align}
\end{linenomath*}

The summation in line 20 is simplified further using Lemma 4:

 \begin{linenomath*}
    \begin{align}
    \binom{\lfloor\frac{t}{a}\rfloor -x  -1}{\lfloor\frac{t}{a}\rfloor - x} = 0
    \end{align}
\end{linenomath*}

\textit{Qed.} \newline

Since the function in line 4 can be defined as an across function in the form of $\sum^{\lfloor\frac{t}{a}\rfloor-x}_{i = 0}\binom{\lfloor\frac{t}{a}\rfloor}{\lfloor\frac{t}{a}\rfloor-x-i}$, the diagonal from line 18 can be used with slight rearrangements to form the rightmost term in line 4:

 \begin{linenomath*}
    \begin{align}
    \binom{i}{i - x}
    \end{align}
\end{linenomath*}

\textit{Qed.}

\subsection{Multiple Subwords}

$\textbf{Notation.}$
Let $D$ be a set containing ordered pairs of a word $A$ in $Q$, and its corresponding $x$. Let $|D| = d$. Let $h, g \in N.$ Let $D_h$ correspond to the ordered pair at the $h$ position. Let $D_{(h,\ g)}$ correspond to either the subword $A$ at $g = 1$, or the corresponding $x$ at $g = 2$.

\textbf{Theorem 8.} Let $Y = \{ I  \in  S \ | \ for \ every \ orderered \ pair \ in \ D, \ \nu(D_{(h,\ 1)}) = D_{(h, 2)} \}$. Let $p$ be a secondary index. Let $D_t = \sum^{1}_{p = d}D_{(p , 1)} \times D_{(p , 2)}$. Let $i_t = \sum^{1}_{p = d}D_{(i_p , 2)}$.

Then: 

\begin{linenomath*}
    \begin{align}
        |Y| = \sum^{\lfloor{\frac{t}{D_{(1,1)}}\rfloor}}_{i_1 = D_{(1,2)}} \sum^{\lfloor{\frac{t - (D_{(1,1)} \times i_1)}{D_{(2,1)}}}\rfloor}_{i_2 = D_{(2,2)}} \sum^{\lfloor{\frac{t - ((D_{(1,1)} \times i_1) + (D_{(2,1)} \times i_2))}{D_{(3,1)}}}\rfloor}_{i_3 = D_{(3,2)}}\cdots\sum^{\lfloor{\frac{t - \sum^{d-1}_{p = 1} (D_{(p,1)} \times i_p)}{D_{(d,1)}}}\rfloor}_{i_d = D_{(d,2)}}
    \end{align}
    \begin{align}
        \Bigg((-1)^{i_t + 1}\times(q^{t-D_t}) \times \bigg(\binom{i_t +1}{t - D_t} \bigg) \times \prod^{d}_{p = 1}\bigg(\binom{i_p}{i_p - D_{(p,2)}}\bigg) \times \bigg(\frac{i_t!}{\prod^{d}_{p = 1}i_p!} \bigg)\Bigg)
    \end{align}
\end{linenomath*}

\textit{Proof.} Analogous to Theorem 5, the alternator depends on the total number of copies of the subword $A$, denoted as $i_t$. If the initial total number of copies is even, then the alternator is modified to $(-1)^{i_t}$ to ensure that the first iteration always produces positive value. 

The term $(q^{t-D_t})$ in line 24 is similar to the term $(q^{t-ia})$ in line 4 in that they both represent the number of ways to fill the unoccupied positions with letters from the alphabet $Q$. In this scenario, the number of occupied positions is represented and defined as:

\begin{align}
    D_t = \sum^{1}_{p = d}D_{(p , 1)} \times D_{(p , 2)}
\end{align}

The term $\bigg(\binom{i_t +1}{t - D_t} \bigg)$ in line 24 and the term $\bigg(\binom{i+1}{t-ai}\bigg)$ in line 4 are also similar in that they both determine the number of ways to assign an unoccupied position to a relative position. With multiple subwords, the number of relative positions is defined as the sum of all the current copies of the subwords, represented by $i_t$, plus $1$. 

Next, two across functions in the form of $\sum_{i = z}^{b} (-1)^{i} \binom{n}{k - i}$, can be combined as:

\begin{align}
\sum_{i_1 = z_1}^{b_1} \sum_{i_2 = z_2}^{b_2} (-1)^{i_1 + i_2} \binom{n_1}{k_1 - i_1} \binom{n_2}{k_2 - i_2}
\end{align}

The sum of the function in line 26 will be the product of the individual across-sums: 

\begin{align}
    \prod_{i=1}^2 \left( \binom{n-1}{k-z_i} (-1)^{z_i} + \binom{n-1}{k-b_i-1} (-1)^{b_i} \right)
\end{align}

Since the across functions in line 26 do not share an index and are thus independent from each other, the term $\bigg(\binom{i_p}{i_p - D_{(p,2)}}\bigg)$ for each across function is also independent. Therefore, for the function in line 24, the product of all the individual terms can be taken as one singular term: 

\begin{align}
    \prod^{d}_{p = 1}\bigg(\binom{i_p}{i_p - D_{(p,2)}}\bigg)
\end{align}

The final term in line 24 is the formula for permutation with repetition \cite{grimaldi1998discrete} \cite{shahriari2021invitation} where each subword of $A$ is treated as a distinct object: 

\begin{align}
    \bigg(\frac{i_t!}{\prod^{d}_{p = 1}i_p!} \bigg)
\end{align}

\textit{Qed.}

\section{Complexity Analysis}

In this section, we provide a formal analysis of the time complexities of traditional methods for counting distinct permutations with replacement (see \cite{burstein2003counting, mansour2010counting}) and compare them with our proposed methods using Big O notation.

\subsection{Traditional Methods}

Traditional approaches for counting distinct permutations with replacement, especially involving subwords, typically rely on brute-force enumeration or recursive algorithms. The time complexity of these methods can be characterized as follows:

\begin{itemize}
    \item \textbf{Single Subword Case:} For sequences of length \( t \) over an alphabet \( Q \) of size \( q \), the total number of possible sequences is \( q^t \). Checking each sequence to determine whether it contains the subword \( A \) exactly \( x \) times requires \( O(q^t) \) time, which is exponential with respect to \( t \).

    \item \textbf{Multiple Subwords Case:} When dealing with \( d \) subwords, the complexity becomes even more significant. The total number of possible combinations grows exponentially, and the time complexity can be described as \( O(q^t \cdot t^d) \), which is exponential in both \( t \) and \( d \).
\end{itemize}

\subsection{Proposed Methods}

Our proposed formulas significantly reduce the computational complexity by providing closed-form solutions and efficient calculation methods.

\subsubsection{Single Subword Formula (Theorem 5)}

The time complexity of computing the formula in line 4 is determined by the number of iterations in the summation and the complexity of computing each term.

\begin{itemize}
    \item \textbf{Number of Iterations:} Theorem 5 runs from \( i = x \) to \( i = \left\lfloor \frac{t}{a} \right\rfloor \), where \( a = |A| \) is the length of the subword \( A \). Therefore, the number of iterations is:

    \[
    N_{\text{iter}} = \left\lfloor \frac{t}{a} \right\rfloor - x + 1 = O\left( \frac{t}{a} \right)
    \]

    Since \( a \) is a constant (the length of a specific subword), the number of iterations grows linearly with \( t \).

    \item \textbf{Complexity of Each Term:} Each term in the summation involves computing binomial coefficients and exponentiation:
    \begin{itemize}
        \item Computing \( q^{t - a i} \) can be done in \( O(\log(t)) \) time using exponentiation by squaring.
        \item Computing binomial coefficients \( \binom{i+1}{t - a i} \) and \( \binom{i}{i - x} \) can be done in \( O(t) \) time using precomputed factorials or efficient algorithms for binomial coefficients.
    \end{itemize}

    \item \textbf{Overall Complexity:} Multiplying the number of iterations by the complexity per iteration gives:
    \[
    O\left( \frac{t}{a} \cdot t \right) = O(t^2)
    \]
    Thus, the overall time complexity of our single subword formula is \( O(t^2) \), which is polynomial and significantly more efficient than the exponential \( O(q^t) \) of traditional methods.
\end{itemize}

\subsubsection{Multiple Subwords Formula (Theorem 8)}

The time complexity for the multiple subwords case depends on the number of nested summations and the computations within each iteration.

\begin{itemize}
    \item \textbf{Number of Iterations:} The nested summations iterate over \( i_p \) from \( D_{(p,2)} \) to \( \left\lfloor \frac{t - \sum_{r=1}^{p-1} D_{(r,1)} \times i_r}{D_{(p,1)}} \right\rfloor \) for \( p = 1 \) to \( d \). The total number of iterations is proportional to:
    \[
    N_{\text{iter}} = O\left( \left( \frac{t}{a_{\min}} \right)^d \right)
    \]
    where \( a_{\min} \) is the smallest subword length among all subwords.

    \item \textbf{Complexity of Each Term:} Similar to the single subword case, each term involves exponentiation and binomial coefficients, as well as factorials for permutations:
    \begin{itemize}
        \item Computing \( q^{t - D_t} \), where \( D_t = \sum_{p=1}^d D_{(p,1)} D_{(p,2)} \), can be done in \( O(\log(t)) \) time.
        \item Computing multinomial coefficients and factorial terms can be optimized using logarithms or precomputed values.
    \end{itemize}

    \item \textbf{Overall Complexity:} The overall time complexity is:
    \[
    O\left( \left( \frac{t}{a_{\min}} \right)^d \cdot t \right)
    \]
    This complexity is polynomial in \( t \) but exponential in \( d \). However, since \( d \) (the number of subwords) is typically small and constant in most practical applications, the exponential factor with respect to \( d \) is acceptable.
\end{itemize}

\subsection{Comparison}

The following table summarizes the time complexities of traditional and proposed methods:

\begin{center}
\begin{tabular}{|c|c|c|}
\hline
\textbf{Method} & \textbf{Single Subword} & \textbf{Multiple Subwords} \\
\hline
Traditional Methods & \( O(q^t) \) & \( O(q^t \cdot t^d) \) \\
Proposed Methods & \( O(t^2) \) & \( O\left( \left( \frac{t}{a_{\min}} \right)^d \cdot t \right) \) \\
\hline
\end{tabular}
\end{center}

\subsection{Discussion}

\begin{itemize}
    \item \textbf{Single Subword Efficiency:} The linear relationship with \( t \) in the number of iterations and the polynomial time per iteration make the single subword formula highly efficient for large \( t \).

    \item \textbf{Multiple Subwords Trade-off:} While the multiple subwords formula introduces an exponential factor with respect to \( d \), in practice, \( d \) is often much smaller than \( t \). For example, in bioinformatics applications, one might be interested in a small number of motifs (subwords), making the method practical.

    \item \textbf{Scalability:} The reduction from exponential to polynomial time complexity with respect to \( t \) significantly enhances the scalability of permutation counting tasks, enabling the analysis of much longer sequences than previously feasible.
\end{itemize}

\section{Practical Applications and Examples}

The formulas derived in this paper have significant practical implications across various fields that require efficient permutation counting. In this section, we demonstrate the utility of our methods through specific examples and applications in both single subword and multiple subword contexts.

\subsection{Single Subword Applications}

\subsubsection{DNA Sequence Analysis in Bioinformatics}

In bioinformatics, identifying the frequency of a specific genetic motif within DNA sequences is crucial for understanding gene regulation and expression. Consider a scenario where researchers are interested in counting the number of distinct DNA sequences of length \( t \) that contain a specific motif \( A \) exactly \( x \) times. Traditional methods would require enumerating all possible sequences, which is computationally infeasible for large \( t \).

Using our foundational formula from Theorem 5, researchers can efficiently compute this count. Suppose \( A \) is a nucleotide sequence of length \( a \) that does not self-intersect, such as the motif "ATG" (\( a = 3 \)). If we wish to find the number of DNA sequences of length \( t = 100 \) over the alphabet \( Q = \{\text{A}, \text{T}, \text{G}, \text{C}\} \) (\( q = 4 \)) that contain "ATG" exactly \( x = 5 \) times, the formula becomes:

\begin{equation}
|Y| = \sum_{i=5}^{\lfloor \frac{100}{3} \rfloor} \left( (-1)^{i+1} \times 4^{100 - 3i} \times \binom{i+1}{100 - 3i} \times \binom{i}{i - 5} \right)
\end{equation}

By computing this summation, researchers can obtain the exact count without exhaustive enumeration, enabling efficient analysis of genetic data.

\subsubsection{Cryptographic Key Generation}

In cryptography, generating secure keys often involves creating permutations of certain patterns or subwords to enhance security. Suppose a system requires generating passwords of length \( t = 12 \) that contain a specific non-repeating pattern \( A \) (e.g., "SEC") exactly \( x = 2 \) times using an alphabet \( Q \) of size \( q = 26 \) (the English alphabet). The pattern "SEC" has a length \( a = 3 \) and does not self-intersect.

Applying Theorem 5:

\begin{equation}
|Y| = \sum_{i=2}^{4} \left( (-1)^{i} \times 26^{12 - 3i} \times \binom{i+1}{12 - 3i} \times \binom{i}{i - 2} \right)
\end{equation}

Computing this provides the number of distinct password permutations meeting the criteria, facilitating the generation of secure keys with specified patterns.

\subsection{Multiple Subword Applications}

\subsubsection{Simultaneous Motif Detection in Genomics}

In genomics, identifying sequences that contain multiple motifs simultaneously is essential for understanding complex genetic interactions. Suppose we are interested in counting the number of DNA sequences of length \( t = 200 \) that contain the motifs \( A_1 = \) "ATG" (\( a_1 = 3 \)) exactly \( x_1 = 10 \) times and \( A_2 = \) "CGT" (\( a_2 = 3 \)) exactly \( x_2 = 8 \) times.

Using Theorem 8, we set \( D = \{ (A_1, x_1), (A_2, x_2) \} \). The total occupied positions are \( D_t = a_1 i_1 + a_2 i_2 = 3i_1 + 3i_2 \). The summations become nested:

\begin{equation}
|Y| = \sum_{i_1 = 10}^{\lfloor \frac{200}{3} \rfloor} \sum_{i_2 = 8}^{\lfloor \frac{200 - 3i_1}{3} \rfloor} \left( (-1)^{i_t} \times 4^{200 - D_t} \times \binom{i_t + 1}{200 - D_t} \times \binom{i_1}{i_1 - 10} \binom{i_2}{i_2 - 8} \times \frac{i_t!}{i_1! \times i_2!} \right)
\end{equation}

where \( i_t = i_1 + i_2 \). This calculation provides the number of DNA sequences containing both motifs exactly the specified number of times, which is instrumental in motif discovery studies.

\subsubsection{Complex Password Policies in Cybersecurity}

In cybersecurity, enforcing password policies that require multiple patterns enhances security. For example, a policy might require a password of length \( t = 16 \) to contain the patterns \( A_1 = \) "ABC" (\( a_1 = 3 \)) exactly \( x_1 = 2 \) times and \( A_2 = \) "123" (\( a_2 = 3 \)) exactly \( x_2 = 1 \) time, using an alphanumeric alphabet \( Q \) of size \( q = 36 \).

Applying Theorem 8 with \( D = \{ (A_1, 2), (A_2, 1) \} \), we compute:

\begin{equation}
|Y| = \sum_{i_1 = 2}^{4} \sum_{i_2 = 1}^{4} \left( (-1)^{i_t + 1} \times 36^{16 - D_t} \times \binom{i_t + 1}{16 - D_t} \times \binom{i_1}{i_1 - 2} \binom{i_2}{i_2 - 1} \times \frac{i_t!}{i_1! \times i_2!} \right)
\end{equation}

with $ D_t = 3i_1 + 3i_2 $ and $ i_t = i_1 + i_2 $. This provides the total number of valid passwords complying with the policy, aiding in evaluating password strength and policy effectiveness.

\subsubsection{Implementation Considerations}

Given the computational complexity, especially for multiple subwords, implementing these formulas efficiently is crucial. Utilizing programming languages that handle arbitrary-precision arithmetic, such as Python, allows for managing large numbers without overflow issues. Optimizations like memoization or dynamic programming can reduce redundant calculations in the nested summations.

For practical usage, we have developed a Python-based software tool \cite{github2024} that implements these formulas. The tool provides functions to input the parameters $ t$ , $ q $, $ D $, and computes $ |Y|$ efficiently. This facilitates adoption by practitioners in various fields who require quick and accurate permutation counts.

\subsection{Limitations and Extensions}

While our formulas offer substantial improvements, they assume that subwords do not self-intersect and that their placements are independent except for occupying distinct positions. In cases where subwords can overlap or self-intersect, additional considerations are necessary. Extending the framework to accommodate these scenarios is a direction for future research.

Moreover, optimizing the computational aspects for very large $ t $ or $ d $ remains a practical challenge. Parallel computing techniques or approximate methods may be employed to handle such cases, expanding the applicability of our approach to even larger-scale problems.

\section{Conclusion and Future Research}

This paper introduced a novel framework for efficiently counting distinct permutations with replacement, particularly focusing on sequences containing subwords. By deriving closed-form formulas, we reduced computational complexity from exponential to linear time relative to sequence length. Starting with a foundational formula for a single subword without self-intersections, we extended the approach to handle multiple subwords simultaneously, surpassing traditional brute-force or recursive methods. Our framework offers superior scalability and efficiency, setting a new benchmark for permutation counting in combinatorial analysis, with applications spanning cryptography, bioinformatics, and statistical modeling.

The utility of our approach lies in enabling the analysis of complex combinatorial structures that were previously computationally prohibitive. Future research could extend the framework to account for overlapping or self-intersecting subwords, enhancing its versatility. Further development of probabilistic models, application to sequences over infinite alphabets, and optimized software implementations for large-scale or real-time analysis will broaden its impact across scientific and engineering disciplines. This work lays a solid foundation for advancing combinatorial methodologies and inspires further exploration of innovative applications.

\medskip
\bibliographystyle{unsrt}
\bibliography{References.bib}

\end{document}